\documentclass[final,authoryear,3p,twocolumn]{elsarticle}
\usepackage{graphicx}
\usepackage{amssymb}
\usepackage{amsmath}
\usepackage{hyperref}
\usepackage{breakurl}

\newcommand{\tss}[1]{$_{\text{#1}}$}

\journal{Atmospheric Environment}

\begin{document}


\begin{frontmatter}

\title{Statistical characterisation of bio-aerosol background in an urban environment}

\author[dsto]{M.~Jamriska\corref{cor1}}
\ead{milan.jamriska@dsto.defence.gov.au}
\author[dsto]{T.C.~DuBois}
\author[dsto]{A.~Skvortsov}

\cortext[cor1]{Corresponding author}
\address[dsto]{Defence Science and Technology Organisation,  506
Lorimer Street, Fishermans Bend, Victoria, 3207, Australia}

\begin{abstract}
In this paper we statistically characterise the bio-aerosol background in an urban environment. To do this we measure concentration levels of naturally occurring microbiological material in the atmosphere over a two month period. Naturally occurring bioaerosols can be considered as noise, as they mask the presence of signals coming from biological material of interest (such as an intentionally released biological agent). Analysis of this `biobackground' was undertaken in the $1$-$10 \mu m$ size range and a $3$-$9$\% contribution was found to be biological in origin - values which are in good agreement with other studies reported in the literature. A model based on the physics of turbulent mixing and dispersion was developed and validated against this analysis. The Gamma distribution (the basis of our model) is shown to comply with the scaling laws of the concentration moments of our data, which enables us to universally characterise both biological and non-biological material in the atmosphere. An application of this model is proposed to build a framework for the development of novel algorithms for bio-aerosol detection and rapid characterisation.
\end{abstract}


\end{frontmatter}

\section{Introduction}
\label{sec:intro}

It is well recognised that biogenic aerosols have a significant effect on human health, the environment, climate, atmosphere and other phenomena (\cite{Poschl05, Jaenicke07, Ho05, Huffman10} and references therein). Numerous studies have been conducted that focus on the characterisation of naturally occurring microbiological material in the atmosphere (referred in this paper as biobackground) (\cite{Bauer08}). This large body of literature indicates a rich and diversified dependence of biobackground characteristics affected by geographical location, temporal, seasonal and daily variations as well as the dependency on applied measuring techniques (\cite{Ho98}). A comprehensive review of publications concerning bioaerosol studies is outside of the scope of the current paper and the readers are referred to a number of recently published reviews (\cite{Huffman10, Burrows09a, Harrison04, Jaenicke07, Ariya09} and references therein).

Biobackground is the primary limiting factor of bio-detection technologies used in operational applications (bio-security, defence systems and counterterrorism) (\cite{Ho05, Jensen07}).

Naturally occurring bioaerosols can be considered as noise, as they mask the presence of signals coming from biological material of interest (e.g.\ an intentionally released biological agent), hindering a biodetectors' capability (by affecting its sensitivity, selectivity, time of response or rate of false alarms) to detect airborne biological threats\footnote{A review of bio-detection methods and technologies is presented for example in \cite{trw01}.}.

Selection of a specific technique for bioaerosol detection is a challenging task, whose outcome is determined by both the intended application and the scenario in which it is deployed (\cite{Buteau10}). It is unlikely that a single biodetection technique capable of detecting all biological agents will be developed in the foreseeable future. This necessitates the use of multiple, often complementary technologies and data processing algorithms for reliable bio-threat detection. The consistent integration of these technologies into a single operational system is influenced by the critical issue of `data fusion' of multiple information sources, i.e.\ the rigorous comparison of the relative information gain that can be provided by a particular bio-detection technique. Apparently this issue cannot be thoroughly investigated and eliminated without detailed knowledge of the biobackground the proposed system is operational in. From this perspective; understanding the characteristics, dependencies and parameters affecting the biobackground is a critical step in the development of enhanced biodetection capability operating over a broad spectrum of applications and scenarios.

Particles in the ambient environment (atmospheric surface layer) are in continuous chaotic motion being advected by turbulent flow. Since most of these particles are very small (less than $10 \mu m$) their inverse effect on flow is negligible and their kinematics can be described by the models of `passive scalars' (i.e.\ a system of passive markers associated with the individual particles randomly moving by turbulence) (\cite{Falkovich01}). Under this condition it is reasonable to assume that the fundamental physical properties of turbulent mixing and not the physical properties of the particles (e.g.\ size, density \emph{etc.}) would strongly emerge in the statistical characteristics of the biobackground. These characteristics are determined by the kinematics of the turbulent flow and are therefore independent of the origin or nature of the particles (i.e.\ biological or non-biological).

An illustrative example of this approach has been recently reported in our publication, \cite{Skvortsovatall:10}, where we established scaling laws of time series for particle concentration and found they comply with specific models of turbulent mixing (\cite{Yee97}, \cite{Villermaux03}). These results provide the motivation and foundation for the current study, i.e.\ the application of the theoretical methods of passive tracer dispersion in the atmospheric surface layer to deduce the universal statistics (scaling laws) of particles regardless of their origin.

The current study focuses on a bio-detection framework which enables a consistent prediction of some of the statistical properties of particle concentrations in the biobackground (moments, probability distribution functions (PDFs), intermittency corrections, \emph{etc.}) by employing well-known properties of turbulent mixing (\cite{Yee97}, \cite{Villermaux03}, \cite{LebedevTuritsyn:04}, \cite{Yee09}, \cite{SkvortsovYee:11}). The statistical framework presented was validated using a large data set of biological and non-biological particle concentrations measured by an optical biodetector utilising laser induced auto-fluorescence characteristics for living organisms (\cite {Hairston97, Ho98,Brosseau00}). We also discuss the applicability of this statistical approach for the development of detection algorithms in operational systems.

The framework was developed in two steps: first, we collected a large data set of physical characteristics of the ambient biobackground. The data were screened, validated, analysed and results compared to literature. This provided the quantification of a validated biobackground data set for an urban/industrial type of ambient environment. We then developed a mathematical model based on statistical properties (such as moments) derived from analysis of time series of measured data, revealing analytical forms of concentration PDFs characteristic of the ambient biobackground. The derived generic PDFs (Probability Distribution Functions) are fundamental to the development of the next generation of biodetection algorithms.

The paper is organized as follows: in the first part we outline the measuring methodology and present quantification results of the biobackground; in the second part we present the statistical model, its validation and the analytical PDF forms derived from our measured data.

\section{Experimental}
\label{sec:methods}
\subsection{Instrumentation}
The atmospheric aerosol was measured and characterised by two instruments: Fluorescent Aerosol Particle System (FLAPS-III, Dycor Technologies) and Aerodynamic Particle Sizer (APS, TSI Model 3321). Both have been used in a variety of aerosol research studies including the areas of environmental and biobackground monitoring (\cite{Huffman10} and references therein), hence only a brief description follows.

FLAPS-III is an integrated optical biodetection system providing near real time (few seconds) point-detection of particles of biological origin (\cite{Dycor1}). It utilises laser induced auto-fluorescence intrinsic to living biological organisms (\cite{Boulet96, Hairston97, Hill99, Ho98, Eng89, Setlow77, Laflamme05, Agranovski05, Huffman10}). The system is comprised of two major components: FLAPS-III (TSI Model 3317) and particle concentrator (XMX/2A, Dycor Technologies). FLAPS-III measures side scattering (SS) and fluorescence (FL) light intensities of a single particle. The fluorescence signal is measured in two wavelength bands corresponding to the emissions from NADH (FL1) and riboflavin (FL2) (\cite{Hairston97, Boulet96, Hill99}). In conjunction with the XMX concentrator the system provides count rates of total (TAP) and fluorescent biological aerosol particles (FBAPs) in the $1$-$10\mu m$ size range. The system does not allow identification or speciation of the detected bio-particles (\cite{Dycor2}). Detailed descriptions of the instrument and operating principles are presented elsewhere (\cite{Hairston97, Pinnick98, Agranovski03}).

FLAPS-III is currently considered as one of the benchmark biodetectors for use in defence applications. The system has been evaluated for accuracy, sensitivity and reliability in numerous laboratory and field studies using biological agents and simulants and has proved to be a reliable and accurate bio-detector (\cite{Ho98, Ho99, Semler04, Agranovski05}). \cite{Huffman10} demonstrated the suitability of FLAPS-III for field studies and biobackground characterisation.

APS is a time-of-flight optical spectrometer measuring \emph{number} concentration and size characteristics of particles in the $0.5$-$20\mu m$ size range. For the purposes of this study the APS measurement data was truncated to the $1$-$10 \mu m$ size range to match data measured by FLAPS-III. Assuming particle sphericity and known density; \emph{surface}, \emph{volume} and \emph{mass} characteristics of particles can be estimated from the measured data. More detail about the APS instrument is presented in (\cite{Baron05,TSI1,Peters03,Tsai04}).

\subsection{Instruments settings and calibration}
APS and FLAPS-III were factory calibrated prior the commencement of the study and maintained during the measuring campaign according to their respective manuals. The operational settings for FLAPS-III recommended by the manufacturer were experimentally validated using aerosol challenges of known characteristics (\cite{Kanaani08, Agranovski04, Huffman10}).

The SS and FL signals measured by FLAPS-III are each gated into overall $32$ bins according to the increasing SS and FL light intensities. Based on experimental results which were in agreement with the manufacturer's recommendation, bins [2-31] (where bin 0 is the first bin) were selected for the measurement of fluorescent (FL1 and FL2) signals, and bins [0-31] for the measurement of SS signals. Settings for LED power and PMT gain were determined using non-fluorescent PSL spheres and NaCl aerosols following the method outlined in Brosseau et al 2000. FLAPS-III operated in a continuous sampling mode with the concentrator flowrate set at maximum ($350 L/min$) and aerosol stream flowrate $1 L/min$.

\subsection{Fluorescent Biological Aerosol Particles} \label{sec:FBAP}
Fluorescent biological aerosol particles (FBAPs) measured by FLAPS-III represent a subset of primary biological aerosol particles (PBAPs). PBAP can be defined as a particle that discernibly is or was all or part of a living organism (\cite{Gabey09, Pinnick09, Huffman10}). \cite {Ho98} demonstrated that the FBAPs measured by FLAPS-III technology are attributed mainly to the presence of biological material. The measured count of FBAPs may be considered as a conservative estimate of the PBAPs' abundance in the ambient air (\cite{Huffman10}).

Some particles, despite being of non-biological origin, also exhibit fluorescence (\cite{Agranovski03b, Sivaprakasam04}), however the contribution of these interferents to the overall FL counts measured by FLAPS-III for most ambient conditions and the size range of interest ($1$-$10 \mu m$) can be neglected (\cite{Pinnick09, Ho99} and references therein).

Since FLAPS-III does not provide direct quantification of aerosol concentration - only an integrated count of particles detected in the sampled volume is available - it was used in parallel with the APS. The concentration of FBAP was then estimated from the total particle concentration measured by APS and the fraction of fluorescent particles in the total count was determined from FLAPS-III data (FL/SS ratio).

\section{Methods}
\label{sec:expmethods}
\subsection{Air Sampling}
The measurements were conducted on the top of a four story building at $12$m height with the instrumentation housed inside an air-conditioned environmental enclosure. The ambient air was sampled through vertical conductive sampling lines protruding about $1.5$m outside of the enclosure's ceiling. Deposition losses of particles in sampling lines were assessed and found to be negligible. Sampling was done on a semi-continuous basis with the interruptions due to instrument maintenance and availability. Both APS and FLAPS-III operated side by side and provided measurements every 5 seconds.

\subsection{Measurement location}
The sampling site was located near Port Melbourne ($-37^\circ 49^\prime 27.45^{\prime\prime}$, $+144^\circ 54^\prime 43.02^{\prime\prime}$), approximately $0.5 km$ South from the Yarra River; $5 km$ West from the Melbourne CBD and $3 km$ North of Port Philip Bay. The site is representative of an urban/industrial environment influenced by marine sources during SE to SW wind conditions. The dominant local sources contributing to the ambient background are attributed to the emissions from several manufacturing factories (car industry; business parks; dockside) in the area and vehicular traffic emissions originating from two major freeways surrounding the monitoring site from East to West (in a semicircle) and to a lesser degree local traffic on nearby roads. The freeways are located about $1$-$2 km$ away from the sampling site and carry, in general, high traffic volume throughout the day with the traffic peak hours between 07:00-10:00 and 16:00-20:00. In terms of local biogenic sources; there is a park area about $1 km$ SSW from the sampling  site. The topography in the area is open and flat.

\subsection{Meteorological conditions}
Representative meteorological data have been obtained from three meteorological stations (Bureau of Meteorology sites) located $5$-$13 km$ North, SW and SE from the sampling site. The air temperature and RH were in the ranges of $1$-$23^\circ$C \& $36$-$100$\% with mean values of $11^\circ$C and $80$\%, respectively. Rainfall for the period was approximately $37 mm$ per month, with an average of $12.5$ rainy days per month. The prevalent wind conditions for the duration of the measuring campaign were WSW with average wind speeds of $2$-$4 m/s$.

\subsection{Data collection and processing}
Data were collected over $28$ days between 27/04/2009 and 23/06/2009. Approximately $500$ hours and $340$ hours of data ($5$ seconds readings) were measured by FLAPS-III and APS, respectively. About $140$ hours of data were measured simultaneously by both instruments. The data obtained were screened for outliers and anomalies, processed and stored in a custom built SQL database. A set of programming tools allowing data manipulation, retrieval and averaging was developed using MATLAB. Exploratory and statistical analysis was performed using MATLAB and S-Plus software packages.

\section{Data Inspection}
\label{sec:results}
Before discussing our turbulent mixing and dispersion model, it is important to investigate some of the properties of the experimental measurements acquired throughout the campaign and validate the data set against similar studies. Simple statistics for APS and FLAPS-III data are presented in Tables \ref{table:stataps} and \ref{table:statFLAPS}, respectively.

\subsection{APS Data}
The median values of particle \emph{number} concentration (dN) and \emph{mass} concentration (dM) in the $1$-$10\mu m$ size range measured by APS were $1.5 \#/cm^{3}$ and $0.006 mg/m^{3}$, respectively. Median values are presented throughout this paper as a better estimate (compared to mean) due to skewed distribution of measured data distribution (\cite{Ho98, Morawska99}).

\begin{table*}[htb]
     \centering
\begin{tabular}{l|cccc}
 Parameter Measured & Mean & STD & Median & IQR \\
\hline
Conc dN [$1$-$10\mu m$] ($\#/cm^{3}$) & 2.189 & 1.970 & 1.500 & 0.840 \\
CMD dN [$1$-$10\mu m$] ($\mu m$) & 1.500 & 0.101 & 1.486 & 0.107 \\
Conc dM [$1$-$10\mu m$] ($mg/m^{3}$) & 0.009 & 0.007 & 0.006 & 0.004 \\
CMD dM [$1$-$10\mu m$] ($\mu m$) & 3.122 & 1.115 & 2.839 & 0.533 \\
\end{tabular}
\caption{\label{table:stataps} Statistics for aerosol data measured by APS ($n\sim245374$; total sampling time $\sim340$ hours). Particle mass concentration dM was estimated from volume concentration dV and a particle density of $\approx 1 g/cm^{3}$. dN and dM denote particle number and mass concentration; CMD denotes particles count median diameter.}
\end{table*}

A relatively low \emph{number} concentration of particles is expected, as the majority of particles in an urban/industrial type of environment is dominated by combustion sources (e.g. traffic emissions); generating predominantly fine, submicrometer particles (\cite{Jamriska08, Mejia07}) which are outside of the size range considered in this study.

Particle \emph{mass} concentration in the $1$-$10\mu m$ size range (dM[$1$-$10\mu m$]) derived from APS \emph{number} data is also relatively low,  reflecting the dominance of fine particles. An indirect comparison of the measured dM results with literature data indicates a good agreement. \cite{Keywood99} characterised PM concentrations in six Australian cities, reporting a mean value of PM\tss{1}/PM\tss{10} ratio of $0.45$; i.e.\ approximately $55$\% of PM\tss{10} \emph{mass} concentration is attributed to particles in the $1$-$10 \mu m$ size range. Using this estimate and an ambient particle density of $1.6 g/cm^{3}$ (\cite{Morawska99,Pitz03}) the estimates for the PM\tss{10} derived from our data are $0.026 mg/m^{3}$ (mean) and $0.017 mg/m^{3}$ (median). These results are in good agreement with the annual mean PM\tss{10} value for Melbourne of $0.023 \pm 0.009 mg/m^{3}$ reported by \cite{Keywood99} as well as EPA data reported for the monitoring time period ($0.020$-$0.024 mg/m^{3}$) (\cite{EPA09}).

In terms of particle size distribution, the corresponding median diameter for  particle \emph{number} and \emph{mass} distribution was $1.5 \mu m$  and $2.8 \mu m$, respectively.  The results presented are based on the analysis of the available data for particles in the $1$-$10 \mu m$ size range. While there is limited information or data on the size distribution statistics of this particular size range, our results are in a relatively good agreement with findings reported by \cite{Huffman10}.

\begin{figure}[htb]
    \includegraphics[width=\columnwidth]{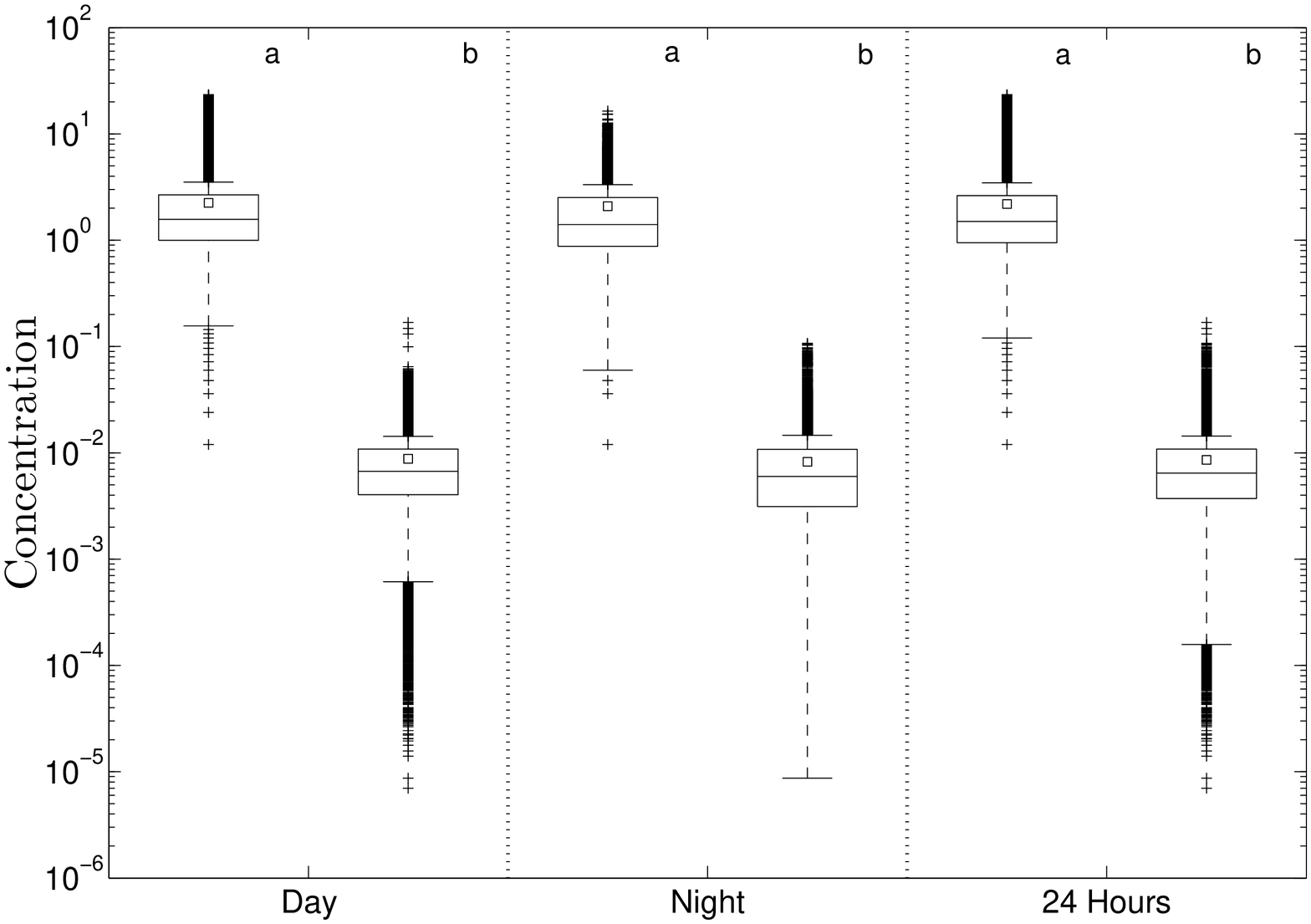}  
    \caption{\label{fig:boxapsdndm}Box plot of particle (a) \emph{number} ($\#/cm^{3}$) and (b) \emph{mass} ($mg/m^{3}$) concentration measured by APS in the size range 1-10$\mu$m during day (06:00-20:00, n$_{Day}\sim 153906 $); night (20:00-06:00), n$_{Night}\sim 91468$); and 24 hour, (n$_{24 hours}\sim 245374$) periods. Mean value displayed as ($\square$); median (centre line); 25\% and 75\% bottom and top of rectangle respectively; whiskers set to $\pm1.93\sigma$ for display purposes; outliers ($+$).} 
\end{figure}

In order to discern emission sources dominating the ambient background during different time periods, the results were grouped into three time intervals: day (06:00-20:00), night (20:00-06:00) and 24h (00:00-24:00) measurements. The statistics for these groups in the form of boxplots is presented in Figure \ref{fig:boxapsdndm}.

Visual observation of the data presented indicates that the mean and median values for the day, night and $24$ hours data sets are comparable for both particle \emph{number} and particle \emph{mass} concentration, however statistical analysis of the medians difference (Mann-Whitney two sample rank-sum test) reveals a significant difference between the medians of all three groups (at the 95\% confidence level). There seems to be less fluctuation (i.e.\ spread of the data as indicated by IQR) in day data as compared to night data. This is likely associated with traffic emissions dominating the ambient background during day time (06:00-20:00) (\cite{Mejia07, BTRE07}).

\begin{figure}[htb]
    \includegraphics[width=\columnwidth]{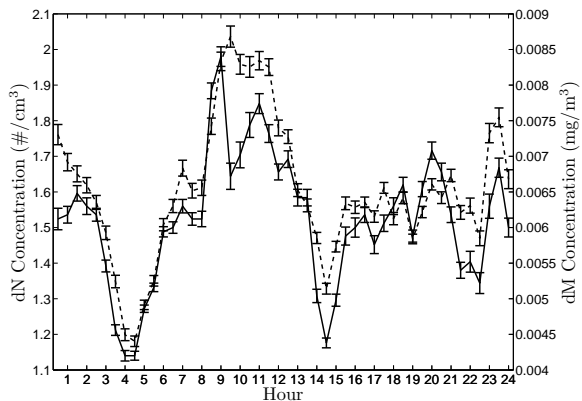}
    \caption{\label{fig:avgdn2medeb110}30 minute average (Median$\,\pm\,IQR/\sqrt{n}$) of \emph{number} (dN, solid line) ($\#/cm^{3}$) and \emph{mass} (dM, dashed line) ($mg/m^{3}$) concentrations measured by APS in the size range 1-10$\mu$m. (Number of data points $n = 245374$; 5 second readings).} 
\end{figure}

To evaluate further the daily variation in APS \emph{number} and \emph{mass} concentrations, all available data were grouped and averaged over $30$ minute time intervals. The averaged time series presented in Figure \ref{fig:avgdn2medeb110} show a complex multi peak profile with two broad dominant modes (6:00-13:00) and (16:00-21:00) likely to be associated with an increased contribution from traffic during the traffic rush hours. In the morning, maximum number concentration was observed at about 09:00 ($\sim 2 \#/cm^{3}$) while in the afternoon/evening mode the maximum was observed at about 20:00 ($\sim 1.7 \#/cm^{3}$). There seems to be also relatively high concentration at night between 23:00-03:00, which could be associated with the emissions from local industrial sources operating 24/7 (e.g.\ dockside activities).

To investigate the trends further (in an attempt to unmask the effect of local emission sources), the daily averages of particle \emph{number} and \emph{mass} concentration were normalised by the night average (i.e.\ median night values of dN and dM, respectively).

Perhaps not surprisingly, the normalised concentrations show similar trends as observed in Figure \ref{fig:avgdn2medeb110} with two dominant modes observed in the morning and late afternoon/evening. The concentration levels reached up to $1.4$ times the night level between 09:00-11:00 and up to $1.2$ times at around 20:00 (dN). Relatively low concentrations were observed for 03:00-05:00 and 13:00-15:00 time periods. The fluctuations could be associated with changes in emission sources (mainly traffic density) contributing to the measured background and changes in atmospheric conditions driving the aerosol dispersion (wind, convection, \emph{etc.}).

A similar trend as presented in Figure \ref{fig:avgdn2medeb110} was reported in a study conducted in Brisbane by \cite{Mejia07} for a comparable type of sampling environment (emission sources, location and topography). The authors showed a strong dependency of submicrometer particle concentration on traffic emissions (traffic density; type of cars) and wind direction. A review of the dependencies of ambient background on the meteorological conditions is presented for example in \cite{Harrison04}.

\subsection{FLAPS-III Data}
The statistics for FLAPS-III data measured over a period of $500$ hours is presented in Table \ref{table:statFLAPS}. The SS counts represent the total aerosol particles (TAPs) and FL counts the fluorescent biological particles (FBAPs) in the $1$-$10 \mu m$ size range measured over $5$ second sampling intervals. FL1[2-31] and FL2[2-31] correspond to integrated particle count in bins [2-31] for the gated fluorescent light intensity as discussed previously.

\begin{table*}[htb]
     \centering
\begin{tabular}{l|cccc}
 Parameter Measured & Mean & STD & Median & IQR \\
\hline
SS Total Count & 4376 & 4754 & 2849 & 2684 \\
FL1[2-31]/SS Count Ratio & 0.053 & 0.054 & 0.039 & 0.031 \\
FL2[2-31]/SS Count Ratio & 0.072 & 0.071 & 0.051 & 0.044 \\
FL1[2-31]/FL2[2-31] Count Ratio & 0.788 & 0.537 & 0.745 & 0.087 \\
FL1 Total Concentration ($\#/cm^{3}$) & 0.046 & 0.076 & 0.023 & 0.025 \\
FL2 Total Concentration ($\#/cm^{3}$) & 0.062 & 0.108 & 0.030 & 0.033 \\
\end{tabular}
\caption{\label{table:statFLAPS} Statistics for aerosol data measured by FLAPS-III. Analysis of SS Total Count and FL/SS Count Ratios include all data ($N = 359571$; total sampling time $\sim500$ hours) measured in $5$ seconds; Results for FL1 and FL2 Total concentration were derived from FLAPS-III and APS paired data($N = 99999$).}
\end{table*}

The median values for FL1/SS and FL2/SS ratio were $3.9$\% (FL1) and $5.1$\% (FL2), respectively. The mean values were $5.3$\% (FL1) and $7.2$\% (FL2). Correlation analysis of FL1 and FL2 measurements ($5$ second readings) showed high correlation (R$^{2} \sim 99$\%) indicating that fluorescence (from both FL1 and FL2 channels) mostly originated from the same particle. Based on that assumption our results indicate that in terms of particle number, the FBAPs represent about $4$-$5$\% (median) of TAPs.

Correlation between the fluorescent (FL1; FL2) and total (SS) count rate was also high ($0.82 \leq R^{2} \leq 0.84$), which indicates a relatively high association between the FL and SS data.

Fluorescence in channel FL2 (emission wavelength $\sim 580 nm$) was stronger compared to FL1 (emission wavelength $\sim 430 nm$) with the FL1/FL2 count ratio $0.79$ (mean) and $0.75$ (median). The effect could be associated with differences in fluorescence originating from different molecules (e.g.\ NADH and riboflavine) and different organisms (\cite{Agranovski05, Brosseau00}).

The presented results for total (SS count) and FBAPs are relative values; dependent on the physical properties of sampled material, size distribution of aerosols and FLAPS-III sampling characteristics. Using the absolute quantification of particle concentration capability of APS (TAPs), and the method described in Section \ref{sec:FBAP}, concentration estimates from FLAPS-III counts can be calculated. Although the instruments use different measuring techniques (APS: Time-of-flight, FLAPS-III: light scattering), a fairly acceptable correlation value of $69.3$\% between the instruments was achieved over the measuring period.

The measured fluorescent particle concentrations (mean values of $0.046$ (FL1) and $0.062$ (FL2) $\#/cm^{3}$) are in good agreement with the literature. \cite{Huffman10} for comparison, reported the mean number concentration of FBAPs in the $1$-$10 \mu m$ size range of {$0.03$ ($\#/cm^{3}$) - measured in a similar type of ambient environment.

Further exploration of the fluorescent concentration dependencies and behaviour was not the focus of this work and will be presented in a follow up publication. Since the measured FLAPS-III data will be used in relative terms as a ratio of FL/SS or normalised SS count, the terms count and concentration will be used interchangeably throughout the following sections.

Similar to the APS data, the FL/SS ratios derived from FLAPS-III data were analysed for different time intervals (day, night and 24 hours) as presented in Figure \ref{fig:boxCFratioFL12}. Several observations can be made from this data: the FL2/SS values were consistently higher compared to FL1/SS values for both (day and night) datasets; the fluorescent content was higher during the day compared to night ((FL/SS)\tss{Day} $>$ (FL/SS)\tss{Night}); (FL/SS)\tss{Day} values showed a narrower spread (IQR) compared to night results and were shifted towards the lower values (smaller fraction of fluorescence). Analysis of the medians difference (Mann-Whitney two sample rank-sum test) showed statistically significant difference between the medians of FL1/SS and FL2/SS for day, night and 24h time periods (at the 95\% confidence level).

\begin{figure}[htb]
    \includegraphics[width=\columnwidth]{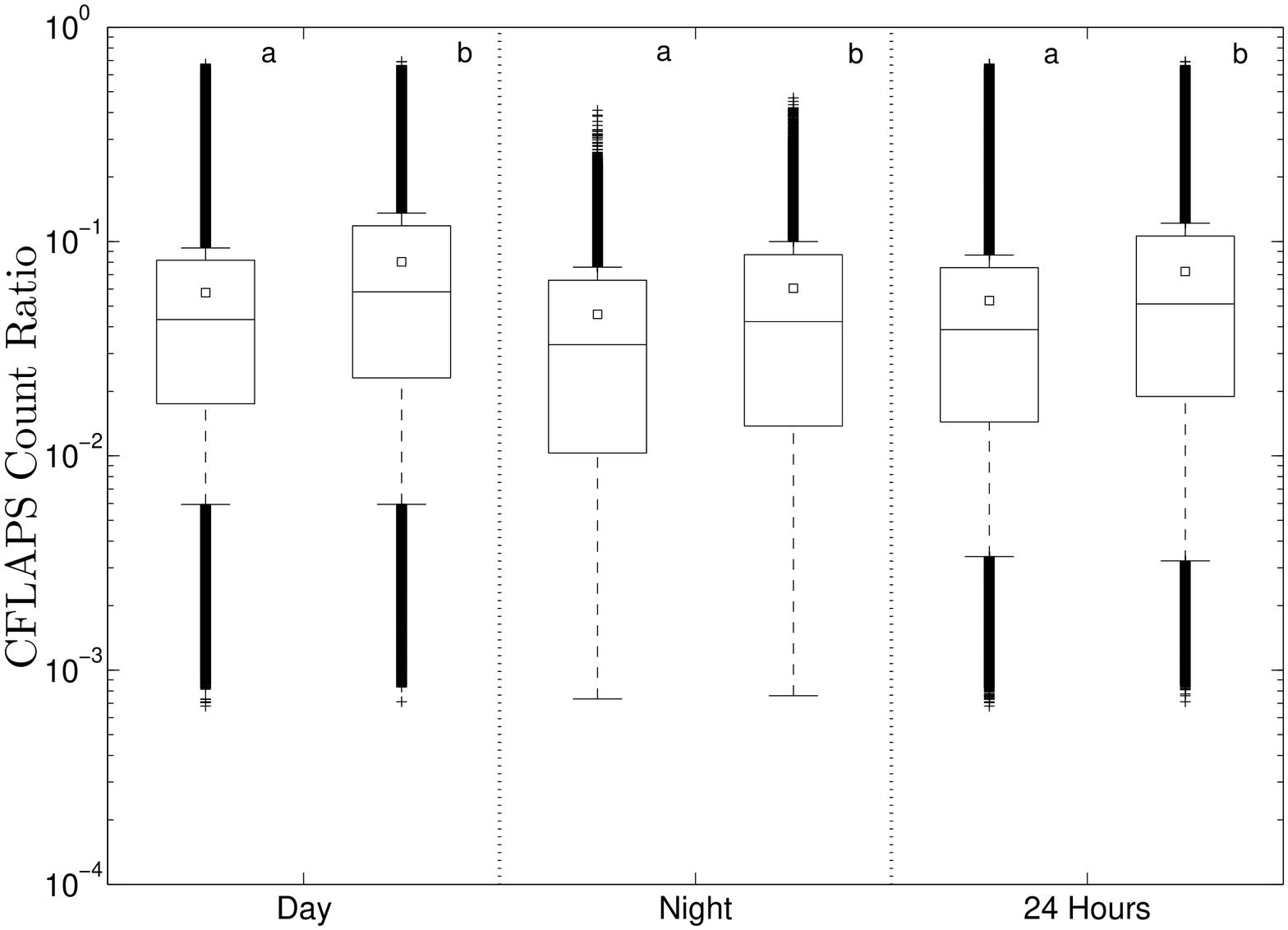} 
    \caption{\label{fig:boxCFratioFL12}Box plot of (a) FL1/SS and (b) FL2/SS count ratio measured by FLAPS-III in the size range 1-10$\mu$m during day (06:00-20:00; N$_{Day}\sim 216188$), night (20:00-06:00; N$_{Night}\sim 143383$) and 24 hour (N$_{24 hours}\sim 359571$) periods. Mean value displayed as ($\square$); median (centre line); 25\% and 75\% bottom and top of rectangle respectively; whiskers set to $\pm1.58\sigma$ for display purposes; outliers ($+$).} 
\end{figure}

The daily variation in fluorescent particle loading calculated as 30 minute averages is presented in Figure \ref{fig:avgcfratio2medeb}. The fraction of fluorescent particles varied throughout the day between $3$-$7$\% (FL1/SS) and $4$-$9$\% (FL2/SS). The variation has diurnal character with the most dominant peaks appearing at around 11:00 and 15:00. The fluorescent particle content showed a steady increase starting at about 5:00 until 9:00, with a sharp increase peaking at 10:30. A second, less pronounced maximum can be observed at about 15:00.

\begin{figure}[htb]
    \includegraphics[width=\columnwidth]{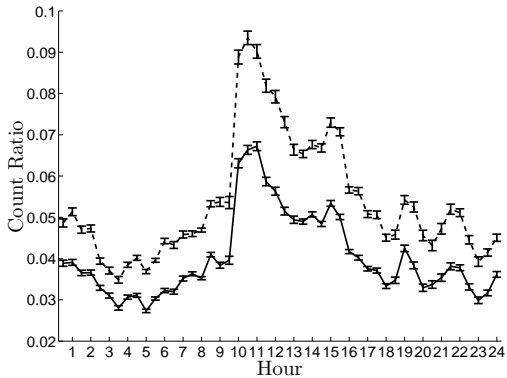}
    \caption{\label{fig:avgcfratio2medeb}FL1/SS (solid line) and FL2/SS (dashed line) Ratio's, 30 minute median averages, $IQR/\sqrt{n}$ shown as error bars.}
\end{figure}

To discern the dependencies and behavior of the 24h daily data, the daily variation of FL and SS data were normalised by the night values as presented in Figure \ref{fig:avgcfnorm2med}. It can be seen that the SS data shows several peaks throughout the day (the most dominant being between 10:00-12:00) while FL1 and FL2 signals exhibit two broad distinct modes (peaks). Similar to the SS data, the most dominant peaks for FL1 and FL2 are observed at about 10:00. The FL levels start to increase at about 05:00 which could be associated with the emission from biogenic sources linked to sunrise with a further steady increase between 08:00-09:00 and then a very rapid climax with a peak at 10:30. We speculate that the dominant mode during 10:00-12:00 period could be associated with the presence of biological material originating from different biogenic sources (e.g. marine environment) emerging as a result of changed environmental conditions (e.g. wind direction). Other factors influencing FL/SS fluctuations could be attributed to different biological material composition with different strengths of fluorescence (\cite{Coz10}). Investigation of the composition of the biological materials and their source apportionment was beyond the scope of this study.

\begin{figure}[htb]
    \includegraphics[width=\columnwidth]{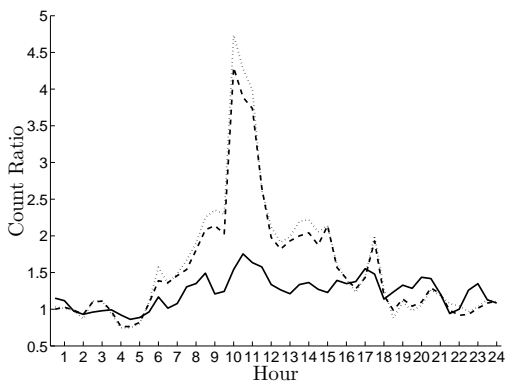}
    \caption{\label{fig:avgcfnorm2med}30 minute averages (Median) for FLAPS-III particle count normalised by the median night count for total (SS [solid line]) and Fluorescent (FL1 [dashed line]; FL2 [dotted line]) particles in the 1-10$\mu$m size range. Median night values of 2430 for SS, 72 for FL1 and 91 for FL2. 359570 data points. 5 second samples.}
\end{figure}

An indirect validation of our results was obtained through comparison with the results of other studies focused on biobackground characterisation using optical and biological methods as outlined below.

\cite{Boulet96}, based on a review of previous studies in different locations, showed that under normal, steady state background atmospheric conditions the fraction of fluorescent particles rarely exceeded $5$\%.

Monitoring of fluorescence in ambient air measured in Calgary CA, where relatively high levels of human and vegetative contributions could be expected, the fluorescent fraction of ambient aerosol in the $2.5$-$10 \mu m$ range rarely exceeded $5$\% regardless of the total particle number concentration (\cite{Ho98}).

\cite{Coz10} reported that PBAPs measured in an urban area in the northeastern Unites States contributed to PM\tss{2.5} on average $6.9 \pm 5.4$\% during summer and $3.3 \pm 1.4$\% during winter. The high variability during summer could be attributed to short term variations in both the load and composition of bacterial aerosols. The small variability found during winter suggest the existence of a relatively constant background level of PM\tss{2.5} PBAPs mass of $2$-$5$\%.

In a study very similar to the presented work, \cite{Huffman10} characterised the biobackground in Mainz, Germany in a semi-urban environment. Using similar experimental and sampling setups (FLAPS-III sampling from a building roof close to the CBD) over several months, the authors reported about $4$\% of FBAPs in terms of particle number and about $20$\% in terms of particle mass.

In summary, the measured data are in a good agreement with results reported in literature in terms of TAP and FBAPs concentration, fluorescent content (FL/SS) and their fluctuation during the day. These results imply validity of the collected data set and its suitability for further exploration using advanced statistical methods aiming to develop superior alarm trigger algorithms.

\section{Model of Biobackground}

Particles in the biobackground are in continuous motion (with random displacements) and are being advected by turbulent flow in the atmospheric surface layer. Due to their size ($1$-$10 \mu m$) the inverse effect of these particles on flow is negligible and their kinematics can be described by the models of `passive scalar' turbulence (or turbulent Lagrangian transport) \cite{Falkovich01}. This means that one can introduce the instantaneous concentration of the particles $\theta(\textbf{r}, t)$ that obeys a stochastic equation of concentration transport (the stochasticity of the transport equation is a result of the randomness of the underlying turbulent velocity field). In other words, the theory of scalar turbulence provides a consistent framework  for  derivation of the statistical properties of concentration  $\theta(\textbf{r},t)$ (i.e.\ moments, PDFs and correlations) for given  statistics of velocity. Remarkably, some important properties of particle statistics experimentally observed in complex environmental settings (boundary layer flow, strong anisotropy, buoyancy \emph{etc.}) emerge from the simplest (but still justifiable) models of turbulent flow i.e.\ white-noise or the Krachnan model (\cite{Falkovich01}). From this perspective we anticipated that certain statistical characteristics of the biobackground should also reflect the fundamental properties of underlying mixing processes, irrespective of the nature (biological/non-biological) of the aerosol particles. This was the main rationale behind our approach to the exploration of our experimental data and the development of a statistical framework as outlined below.

One of the most interesting predictions of the theory of scalar turbulence applied to transport in the atmospheric surface layer is the conclusion that all statistical moments of the scalar concentration $\langle\theta^{n}(\textbf{r},t)\rangle$  are expected to follow the same profile, regardless of the value of $n$. This allows one to write the following simple scaling laws for the concentration moments (for details see \cite{Yee97}, \cite{LebedevTuritsyn:04}, \cite{Yee09}, \cite{SkvortsovYee:11}).
\begin{eqnarray}
\label{e:P1_eq4} \frac{\langle \theta^{n} \rangle }{\langle
\theta\rangle ^{n}} = \xi_n \left (\frac{\langle \theta^{2} \rangle
}{\langle \theta\rangle ^{2}} \right)^{n-1}, ~~ n \geq 2
\end{eqnarray}
and
\begin{eqnarray}
\label{e:P1_eq5} {\langle \theta^{2} \rangle } \propto { \langle \theta\rangle
} , ~~ n = 2.
\end{eqnarray}

The pre-factor  $\xi_n$ in this expression has been theoretically evaluated for two extreme cases. For the case of a perfectly mixed tracer (which usually corresponds to locations far away from a tracer source) $\xi^{-}_n = 1$; while for the opposite, highly intermittent case (near the source region) $\xi^{+}_n  = {\Gamma(n+1)}/{2^{n-1}}$; where $\Gamma(\cdot)$ is the Gamma function (\cite{SkvortsovYee:11}). Hence, in real situations we expect that particle concentration statistics should follow the scaling law ($\ref{e:P1_eq4}$) with the pre-factor $\xi_n$ located between these two estimates ($\xi^{-}_n \leq  \xi_n \leq \xi^{+}_n$).

We found these predictions are in good agreement with our extensive observational data (see Figures \ref{fig:ngt2} and \ref{fig:ne2}). The statistical moments of particle concentrations $\langle\theta^{n}(\textbf{r},t)\rangle$ were calculated from time series recorded by the FLAPS-III (SS, FL1, FL2) and APS instruments at the same location, so $\theta(\textbf{r},t) \equiv \theta(t)$ in our case. The scaling laws (\ref{e:P1_eq4}) corresponding to the experimental data from FLAPS-III for $n = \{3, 4\}$ are depicted in Figure \ref{fig:ngt2} and compared with the theoretical predictions (\ref{e:P1_eq4}) and (\ref{e:P1_eq5}).

The non-biological contribution (top plot in Figure \ref{fig:ngt2}) comes from the SS channel of FLAPS-III and the biological (bottom plot in Figure \ref{fig:ngt2}) from the summation of the FL1 and FL2 channels. Since FL1 and FL2 are strongly correlated the aggregated FL1+FL2 metric captures all relevant data and removes the need to analyse both channels separately. APS concentrations follow the same trends and are therefore not shown in the following analysis.

The experimental data follows the scaling laws (\ref{e:P1_eq4}) reasonably well within a wide range of measured concentrations (see Figure \ref{fig:boxCFratioFL12} and Table \ref{table:statFLAPS}) with the pre-factor value $\xi_n$ extensively occurring in the expected range $\xi^{-}_n \leq  \xi_n \leq \xi^{+}_n$. For the $n = 2$ case (\ref{e:P1_eq5}), a similar plot is presented in Figure \ref{fig:ne2}. Since pre-factor $\xi_2$ is not defined in (\ref{e:P1_eq5}), a value of $3.5$ was assigned for better visual appearance on the theoretical fit. From Figure \ref{fig:ngt2}, we can see that when applied to the biobackground the scaling laws (\ref{e:P1_eq4}), (\ref{e:P1_eq5}) become less accurate for higher order moments (i.e.\ higher $n$) from the presence of outliers above the $\xi^{+}_4$ bound on the bottom plot of Figure \ref{fig:ngt2}. This is similar to the case of tracer dispersion from localised sources (\cite{YeeSkvortsov:11}). Nevertheless, with the two theoretical values of pre-factor they always provide a valuable tool for the estimation of the statistical moments of the aerosol particle concentration in the atmospheric surface layer.

\begin{figure}[ht!b] 
\includegraphics[width=\columnwidth]{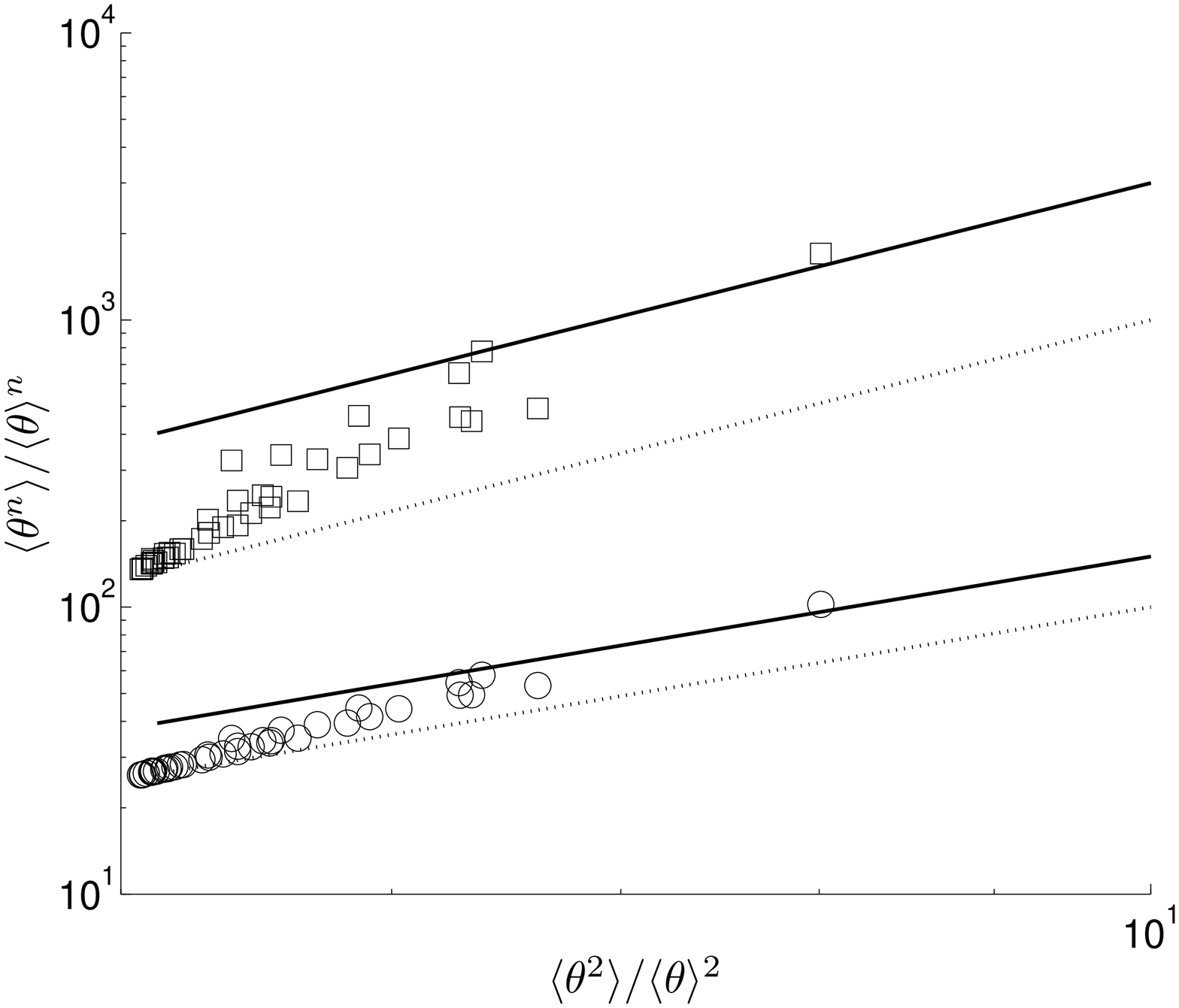}
 \includegraphics[width=\columnwidth]{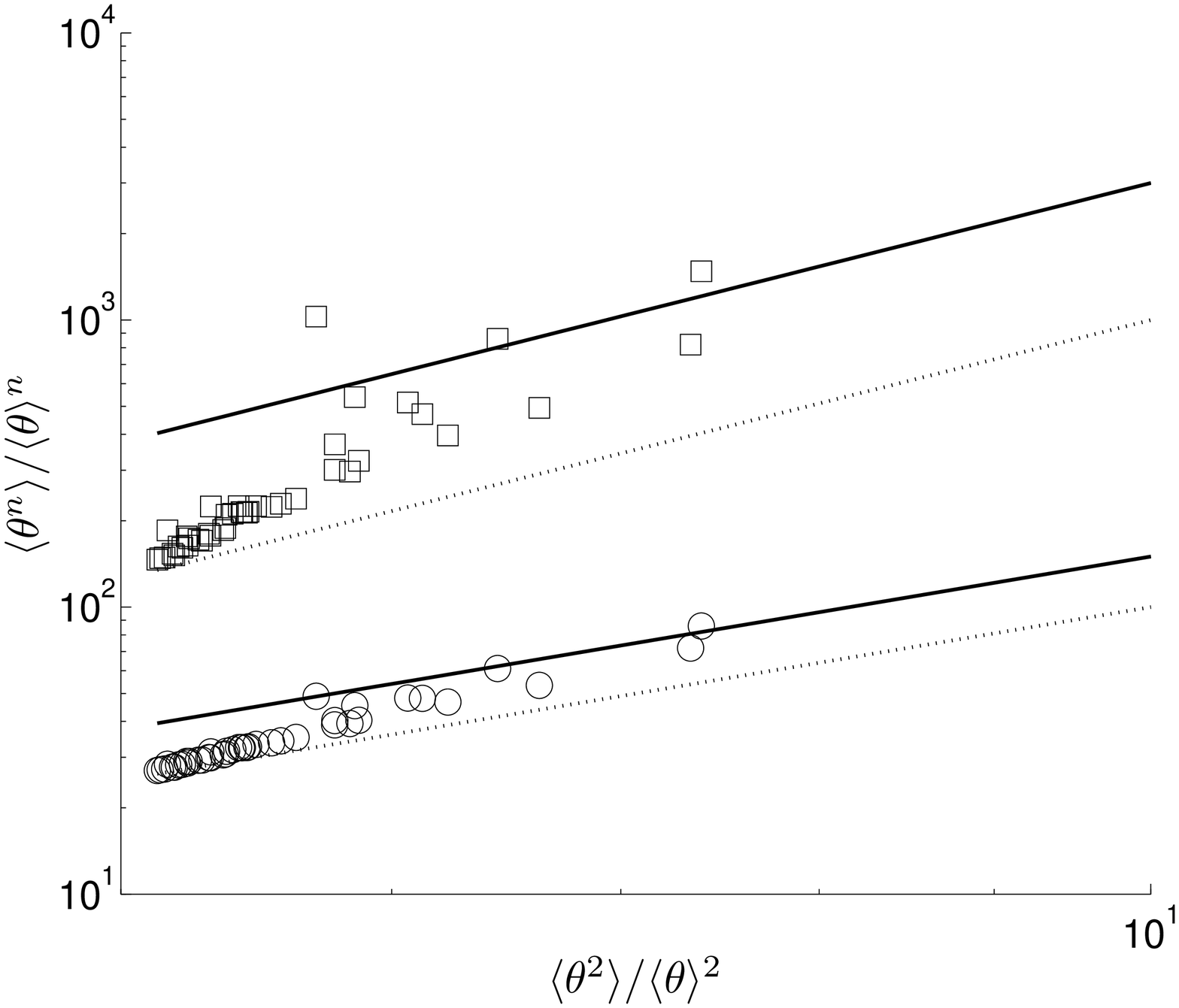}
    \caption{\label{fig:ngt2}Scaling laws for statistical moments of particle concentration. Top plot represents non-biological material (SS) and the bottom biological (FL1+FL2). Bottom and top data series on each plot correspond to moments $n=3$ ($\circ$) and $n=4$ ($\square$) respectively. Lines correspond to the theoretical relationship (\ref{e:P1_eq4})
    with two different values of pre-factor $\xi_n = \xi^{-}_n$ (dashed line),  $\xi_n = \xi^{+}_n$ (solid line), see text for details.}
\end{figure}

\begin{figure}[htb]
\includegraphics [width=\columnwidth]{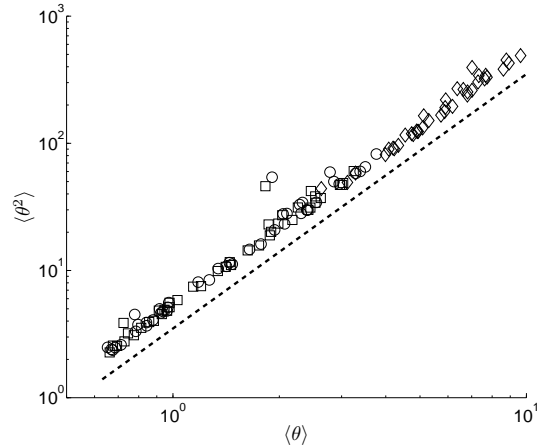}
    \caption{\label{fig:ne2}Similarity of scaling laws for statistical moments of particle concentration for biological (FL1: $\square$, FL2: $\circ$) and non-biological (SS: $\lozenge$) particles, see text for details. The dashed line corresponds to the theoretical prediction (\ref{e:P1_eq5}), with a scaling constant of $3.5$ for visual appearance only.}
\end{figure}

To validate our scaling conjecture further and to prove its independency of the particles' origin we present the scaling law (\ref{e:P1_eq5}) of biological and non-biological sources in Figure \ref{fig:ne2}. We observe that the scaling law (\ref{e:P1_eq5}) of the statistical moments of the biobackground concentration are in excellent agreement with the prediction of the scalar turbulence theory, i.e.\ they are mostly determined by the kinematics of the turbulent surface layer.

It is worth stressing that the scaling laws (\ref{e:P1_eq4}) and (\ref{e:P1_eq5}) manifest a general property of turbulent mixing and are not a consequence of any particular functional form of concentration PDF $p(\theta)$. Contrarily, any distinct PDF form should comply with these scaling laws and this compliance is one of the important criterion for the evaluation of appropriate candidate model for $p(\theta)$.

Amongst the different functional PDF forms of tracer concentration evaluated in studies of turbulent tracer dispersion, the Gamma form has recently emerged as one that adequately captures the physics of turbulent transport and provides a good match with a variety of experimental data (\cite{Yee97, SkvortsovYee:11}). This model (see the Gamma form of (\ref{e:P1_eq42a})) can be analytically derived  using a phenomenological analogy between the mixing and convolution processes by explicit modeling of the deformation of tracer blobs by random stretching and folding (\cite{Duplat08, Villermaux03,Venaille08}). Indeed, the Gamma PDF complies with the scaling laws of the concentration moments (\ref{e:P1_eq4}),(\ref{e:P1_eq5}) with the pre-factor $\xi_n$ residing within the range $\xi^{-}_n \leq  \xi_n \leq \xi^{+}_n$ described above (for details see \cite{YeeSkvortsov:11}). Whilst the Gamma distribution is defined with two parameters (scale $\theta$ and shape $k$), it can also be written in a ``single-parameter'' form (\cite{Yee97, Yee09, Duplat08, Villermaux03}):
\begin{eqnarray}
\label{e:P1_eq42a}
p\bigl(\raisebox{.2ex}{$\chi$} \equiv \theta/\langle \theta \rangle\bigr) = \frac{k^k}{\Gamma(k)} \raisebox{.2ex}{$\chi$}^{k-1} \exp (- k \raisebox{.2ex}{$\chi$}),
\end{eqnarray}
where $k$ is the only fitting parameter of the model and can be related to the statistical moments by (\cite{YeeSkvortsov:11})
\begin{eqnarray}
\label{e:kmom}
\frac{1}{k}=\frac{\langle \theta^{2} \rangle}{\langle \theta\rangle ^{2}}-1.
\end{eqnarray}

From these two equations we can see that our model of the biobackground is uniquely determined by the first two moments (i.e.\ $\langle \theta \rangle$ and $\langle \theta^{2} \rangle$), which is a well known property of the Gamma distribution.

\begin{figure}[htb]
 \includegraphics[width=\columnwidth]{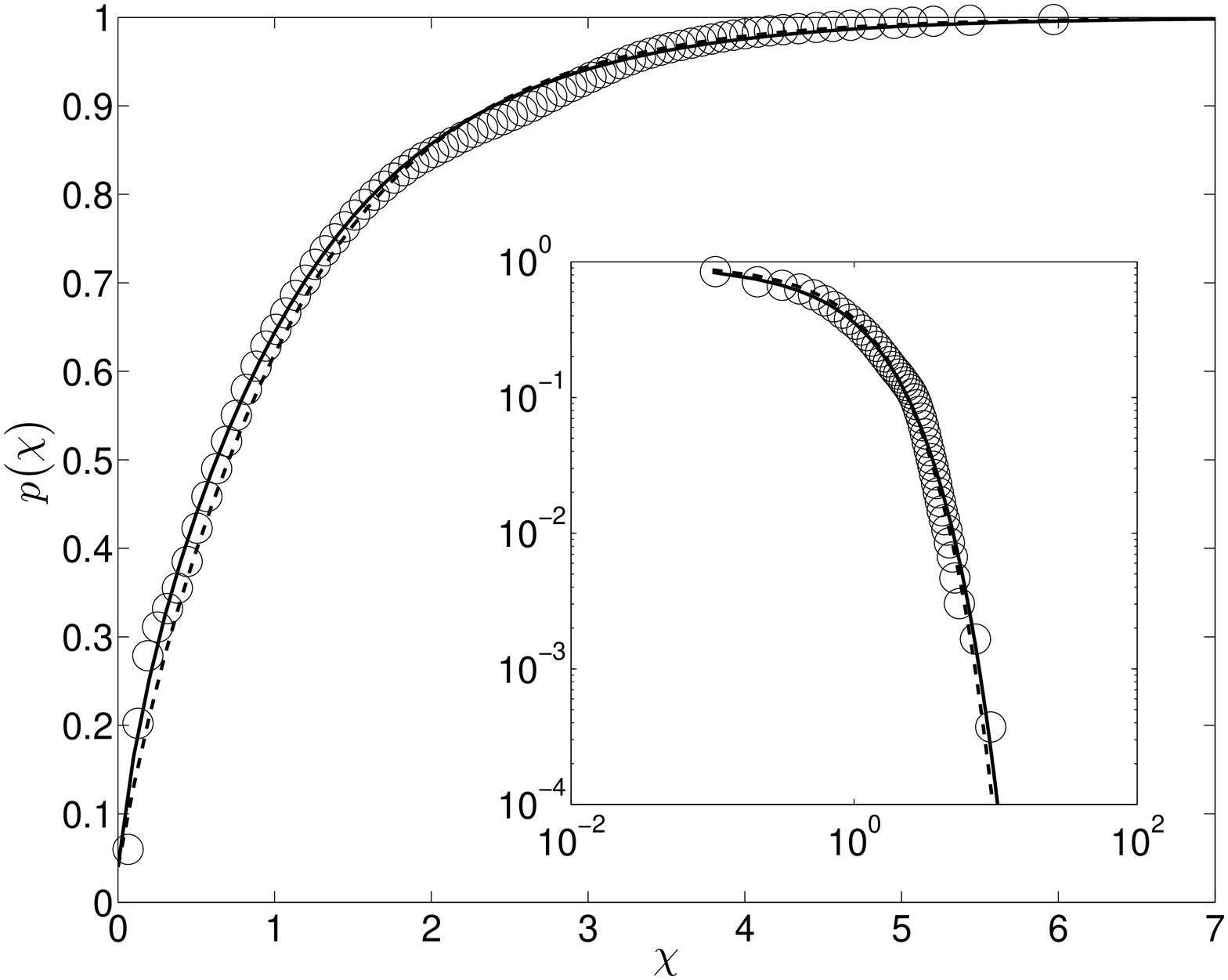}
    \includegraphics[width=\columnwidth]{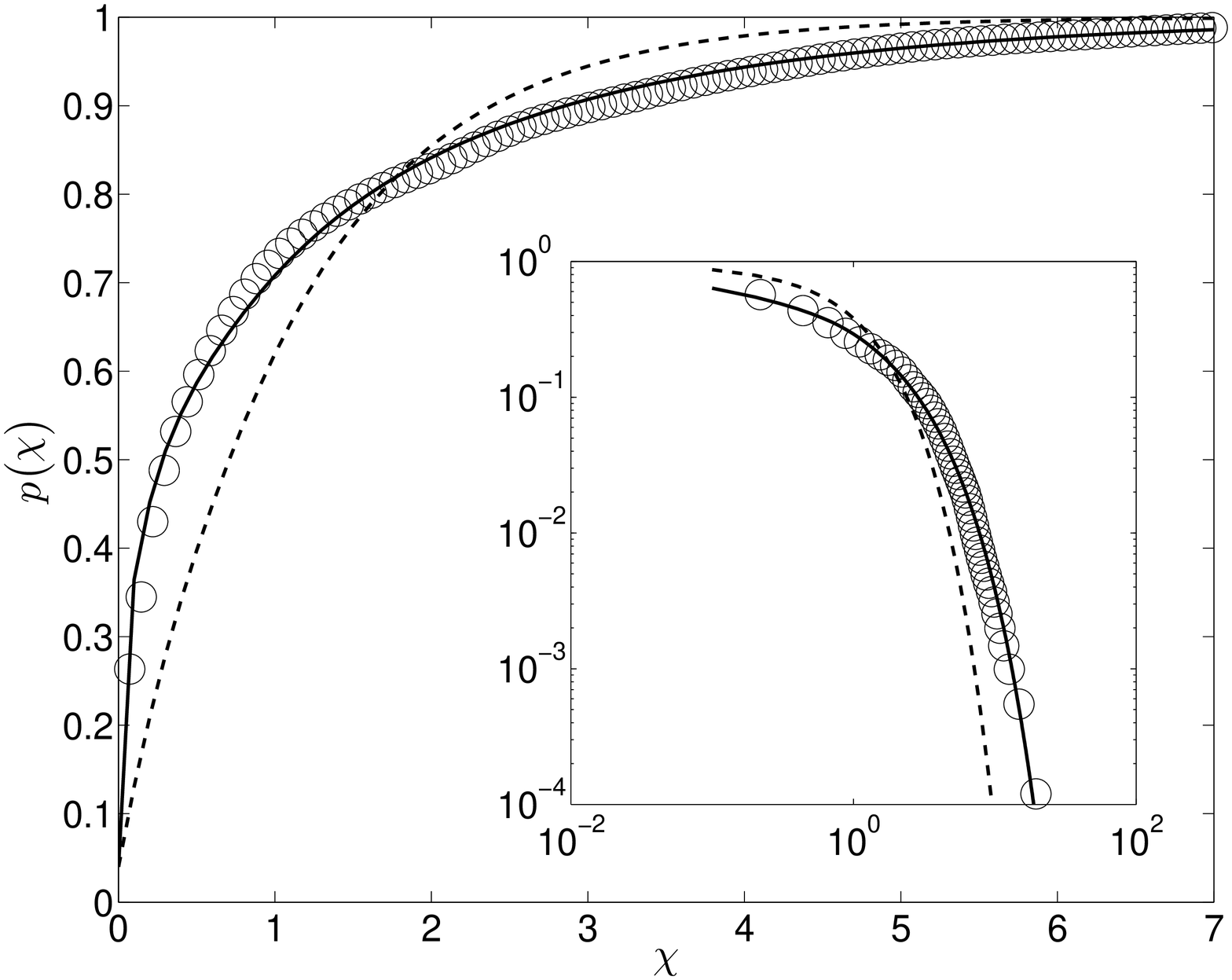}
    \caption{\label{fig:cdfs}CDFs of non-biological (SS, top plot) and biological (FL1+FL2, bottom plot) particle concentrations ($\circ$) with intermittency correction (\ref{e:P1_eq42b}). Gamma (solid line) and Exponential (dashed line) models are also shown. Inserts show the Exceedance Distribution Function (1-CDF) of the same data and models, see text for details.}
\end{figure}

To verify the Gamma PDF model agrees with our experimental data, we compare it with the statistical properties of our observations (i.e.\ time series of particle concentrations from FLAPS-III). In order to do this; $\langle\theta\rangle$ is calculated for the bio- and non bio- particles, then histograms of our time series of concentration in terms of \raisebox{.2ex}{$\chi$}$\: \equiv \theta/\langle \theta \rangle$ were produced. These histograms (or empirical Cumulative Distribution Functions (CDFs)) are compared with the CDF corresponding to the functional form of (\ref{e:P1_eq42a}). In such an approach, $k$ was the only parameter to fit from observations.

The results of this comparison are depicted in Figure \ref{fig:cdfs} (non-biological particles top; biological bottom). The tight fit shows that this model can be reliably used for the parametrisation of the statistical properties of ambient air (both biological and non-biological contributions). The estimations of parameter $k$ were quite stable leading to values $k = 1.1$ for non-biological and $k = 1.8$ for biological particles. For non-biological particles the (mathematically) simpler Exponential PDF (i.e.\ (\ref{e:P1_eq42a}) with $k=1$) can be used as an approximate model for our data. This distribution has a $k$ value of $1.1$ and therefore a relatively small deviation of $0.1$ from the $k=1$ required for the Exponential form. The significant advantage of the Exponential model (even taking into account its inferior data fit compared to Gamma, see both plots on Figure \ref{fig:cdfs}) is that it requires fewer parameters to estimate and to fit. For instance; the Exponential PDF model can be used for rapid calibration of a detection system when only a limited set of data for the monitored environment has been collected. The biological contribution (bottom plot of Figure \ref{fig:cdfs}) is described quite poorly by the Exponential PDF. This contribution (the presence of biological material) can therefore be easily identified just by monitoring the value of $k$. Investigation of this dependency will be the first parameter of interest in our future publication on biodetection algorithms.

The Gamma model (\ref{e:P1_eq42a}) does not take into account the intermittency of a particle distribution (i.e.\ a finite probability that the particle concentration is exactly zero for a certain period of time). This model is therefore only valid for perfectly mixing tracer flows - which rarely occur naturally. In order to estimate the effect of intermittency we appropriately modified our concentration PDFs by adding an intermittency correction (for details see \cite{Yee97, Yee09, SkvortsovYee:11} and references therein)
\begin{eqnarray}
\label{e:P1_eq42b}
p\bigl(\theta\bigr) = (1-\gamma)\delta(\theta) + \gamma \widetilde{p}\bigl(\theta\bigr),
\end{eqnarray}
where $\delta(\cdot)$ is the delta function, $\gamma$ is the intermittency factor (a degree of mixing with $\gamma = 1$ corresponding to the perfect mixing case), $\widetilde{p}\bigl(\theta\bigr)$ is the ``regular'' PDF (in our case (\ref{e:P1_eq42a})). The estimation of parameter $\gamma$ in the model (\ref{e:P1_eq42b})  from concentration time series  is well established in the studies of turbulent dispersion (\cite{Yee97, Yee09, SkvortsovYee:11} and references therein), so we omit all technical details and present only the main results. We found that the effect of $\gamma$ is not significant in our data, i.e.\ $\gamma$ was always close to unity. A stable estimation from both FLAPS-III and APS yield a consistent value $\gamma \approx 0.97$, which is exemplified in the small positive shift in the $p\bigl($\raisebox{.2ex}{$\chi$}$\bigr)$-intercept from $0$ in Figure \ref{fig:cdfs}. Nevertheless, since the value of $\gamma$ is also determined by the proximity to a particle's source, we expect that in some cases the intermittency correction in the model (\ref{e:P1_eq42b}) may become appreciable (\cite{Yee97, Yee09}).

If this is the case, $\gamma$ will be the second parameter of interest in our biodetection algorithm. This assumption also leads to two important conclusions. Firstly, there should be a significant similarity between bio- and non bio-aerosol characteristics measured at the same location (since they undergo the same mixing process). Secondly, these characteristics should clearly comply with the physics of turbulent mixing. As illustrative examples we mention here a universal relationship between statistical moments of particle concentration (a so-called scaling law) and the explicit functional form for the probability distribution function of particle concentration (the Gamma function). It is worth remarking in this context, that the scaling law for statistical moments is more general and is not bound to a particular functional form of concentration PDF (\cite{LebedevTuritsyn:04}).

\section{Conclusions and Future Work}
\label{sec:conclude}

In summary, we quantified the content of ambient background for an urban/industrial type of environment close to the Melbourne CBD. Concentration of total and biological aerosols in the $1$-$10 \mu m$ size range was monitored over a period of two months and the results of exploratory analyses for day, night and daily ($24$ hour) fluctuations are presented.

The measures of total particle \emph{number} and \emph{mass} concentration are in good agreement with previous results obtained for the Melbourne area. Daily profiles ($30$ minute averages) shows complex behaviour with two dominant modes observed in the morning and afternoon/evening corresponding to traffic peaks hours. The results indicate that the ambient background loading is dominated by vehicular traffic emissions.

The fraction of fluorescent particles in total was in the range of $3$-$9$\% (30 minute average) which is in good agreement with other studies reported in the literature. Concentration levels of biological material showed diurnal variation which could be associated with an increase in emissions and changes of biogenic sources. The levels increased up to four times during the midday period as compared to the night mean values. Our results are in good agreement with the study \cite{Huffman10}; conducted in a similar type of ambient environment.

A universal model of the biobackground has been produced from this data set that provides a rigorous framework for the development of novel algorithms for bio-aerosol detection/characterisation and can help to both improve and optimise some existing algorithms used in operational systems.

These refinements may be implemented by initially using the scaling laws (\ref{e:P1_eq4}) and (\ref{e:P1_eq5}), which provide an efficient way to estimate any statistical moment from a limited data set. For instance, moments can be obtained even if only a few concentration levels are known (measured) or measurements are of poor quality. This implies that by employing relationship (\ref{e:P1_eq4}) we can continuously infer the mixing  state of the biobackground, i.e.\ how close its distribution is to the equilibrium state. In effect, any deviation from the equilibrium scaling given by law (\ref{e:P1_eq4}) would indicate the proximity of a bio-aerosol release. Secondly, the specific PDF form the biobackground (i.e.\ Gamma PDF) enables a rigorous assignment of a well-defined measure (based on likelihood or risk) that is associated with any detected spike in bio-aerosol measurement. An informative decision regarding the relevance of this spike can then be made according to the current operational context (e.g.\ it may provide a consistent methodology to reduce the false alarm rate), resulting in rigorous risk mitigating strategies. Lastly, the deep analogy between bio- and non-bio particle distributions emerging from the physics of turbulent mixing allows significant simplification of any calibration process for deployable systems (including the capability for rapid self-calibration), by extracting additional statistical information from monitoring the non-biological aerosol background.

A monitoring system such as this would enable conscious operational decisions to be employed which can help to develop more robust instrumentation algorithms for the detection of biothreats in operational settings, rigorously evaluate the  risk associated with miss and false detection as well as to propose relevant mitigation strategies in response to the detected bio-release. Specifics concerning this framework will be elaborated upon in future publications.

\section{Acknowledgements}
\label{sec:Acknowledge}

This work was supported by DSTO Task 07/086 (Civilian Counter Terrorism and National Security-Public Safety Program) and DSTO Task 07/301 (BioCERP, Bio-surveillance).

The authors would like to acknowledge assistance and support received from Dinesh Pitaliadda, Mark Ryan, Andrew Walker, Gerry Parkinson, Wendy Muir and other colleagues from DSTO who contributed to the project. We thank Chris Woodruff for his careful reading of the manuscript and helpful comments.

\section{References}
\label{sec:Reference}

\bibliographystyle{model2-names}
\bibliography{biobackground}

\end{document}